\documentclass[preprint,aps,letterpaper]{revtex4}
\usepackage[hmargin=1.0in,vmargin=1.0in]{geometry}

\begin{document}
\title{Quantum Simulators: Architectures and Opportunities}

\author{Ehud Altman$^1$}
\author{Kenneth R. Brown$^2$}
\author{Giuseppe Carleo$^3$}
\author{Lincoln D. Carr$^{4*}$}
\author{Eugene Demler$^5$}
\author{Cheng Chin$^6$}
\author{Brian DeMarco$^7$}
\author{Sophia E. Economou$^8$}
\author{Mark A. Eriksson$^9$}
\author{Kai-Mei C. Fu$^{10}$}
\author{Markus Greiner$^5$}
\author{Kaden R. A. Hazzard$^{11}$}
\author{Randall G. Hulet$^{11}$}
\author{Alicia J. Koll\'{a}r$^{12}$}
\author{Benjamin L. Lev$^{13}$}
\author{Mikhail D. Lukin$^5$}
\author{Ruichao Ma$^{14}$}
\author{Xiao Mi$^{15}$}
\author{Shashank Misra$^{16}$}
\author{Christopher Monroe$^{12}$}
\author{Kater Murch$^{17}$}
\author{Zaira Nazario$^{18}$}
\author{Kang-Kuen Ni$^{19}$}
\author{Andrew C. Potter$^{20}$}
\author{Pedram Roushan$^{15}$}
\author{Mark Saffman$^9$}
\author{Monika Schleier-Smith$^{13}$}
\author{Irfan Siddiqi$^1$}
\author{Raymond Simmonds$^{21}$}
\author{Meenakshi Singh$^{4}$}
\author{I.~B.~Spielman$^{12}$}
\author{Kristan Temme$^{18}$}
\author{David S. Weiss$^{22}$}
\author{Jelena Vu\v{c}kovi\'{c}$^{23}$}
\author{Vladan Vuleti\'{c}$^{24}$}
\author{Jun Ye$^{25}$}
\author{Martin Zwierlein$^{24}$}

\affiliation{$^1$Department of Physics, University of California, Berkeley, CA, 94720}
\affiliation{$^2$Department of Electrical and Computer Engineering, Department of Physics, and Department of Chemistry, Duke University, Durham, NC, 27708}
\affiliation{$^3$Center for Computational Quantum Physics, Flatiron Institute, New York, NY 10010}
\affiliation{$^4$Department of Physics, Colorado School of Mines, Golden, CO, 80401}
\affiliation{$^5$Department of Physics, Harvard University, Cambridge, MA, 02138}
\affiliation{$^6$Department of Physics, Chicago University, Chicago, IL, 60637}
\affiliation{$^7$Department of Physics and IQUIST, University of Illinois, Urbana-Champaign, IL, 61801}
\affiliation{$^8$Department of Physics, Virgina Tech, Blacksburg, VA, 24061}
\affiliation{$^9$Department of Physics, University of Wisconsin-Madison, Madison, WI, 53706}
\affiliation{$^{10}$Department of Electrical and Computer Engineering and Department of Physics, University of Washington, Seattle, WA, 98195}
\affiliation{$^{11}$Department of Physics and Astronomy, Rice University, Houston, TX, 77005}
\affiliation{$^{12}$Joint  Quantum  Institute and Department  of  Physics, University  of  Maryland,  College  Park,  MD,  20742, and IonQ, Inc., College Park, MD  20740}
\affiliation{$^{13}$Department of Physics and Department of Applied Physics, Stanford University, Stanford, CA, 94305}
\affiliation{$^{14}$Department of Physics and Astronomy, Purdue University, West Lafayette, IN, 47907}
\affiliation{$^{15}$Google, Inc., Santa Barbara, CA, 93101}
\affiliation{$^{16}$Sandia National Laboratory, Albuquerque, NM, 87123}
\affiliation{$^{17}$Department of Physics, Washington University, St. Louis, MO, 63130}
\affiliation{$^{18}$IBM T.J. Watson Research Center, Yorktown Heights, NY, 10598}
\affiliation{$^{19}$Department of Chemistry and Chemical Biology, Harvard University, Cambridge, MA, 02138}
\affiliation{$^{20}$Department of Physics, The University of Texas at Austin, Austin, Texas, 78712}
\affiliation{$^{21}$National Institute of Standards and Technology, Boulder, CO, 80305}
\affiliation{$^{22}$Department of Physics, The Pennsylvania State University, University Park, PA, 16802}
\affiliation{$^{23}$Department of Electrical Engineering and Ginzton Laboratory, Stanford University, Stanford, CA, 94305}
\affiliation{$^{24}$Center for Ultracold Atoms, Research Laboratory of Electronics, and Department of Physics, Massachusetts Institute of Technology, Cambridge, MA, 02139}
\affiliation{$^{25}$JILA, National Institute of Standards and Technology and University of Colorado, Boulder, CO, 80309}
\affiliation{*To whom correspondence should be addressed: lcarr@mines.edu}

\begin{abstract}
Quantum simulators are a promising technology on the spectrum of quantum devices from specialized quantum experiments to universal quantum computers.  These quantum devices utilize entanglement and many-particle behaviors to explore and solve hard scientific, engineering, and computational problems.  Rapid development over the last two decades has produced more than 300 quantum simulators in operation worldwide using a wide variety of experimental platforms.  Recent advances in several physical architectures promise a golden age of quantum simulators ranging from highly optimized special purpose simulators to flexible programmable devices. These developments have  enabled  a convergence of ideas drawn from fundamental physics, computer science, and device engineering.  They have strong  potential  to address problems of societal importance, ranging from understanding vital chemical processes,  to  enabling the design of new  materials with enhanced performance, to solving complex computational problems. It is the position of the community, as represented by participants of the NSF workshop on ``Programmable Quantum Simulators,'' that investment in a national quantum simulator program is a high priority in order to accelerate the progress in this field and to result in the first practical applications of quantum machines.   Such a program should address two areas of emphasis: (1) support for creating quantum simulator prototypes usable by the broader scientific community, complementary to the present universal quantum computer effort in industry; and (2) support for fundamental research carried out by a blend of multi-investigator, multi-disciplinary collaborations with  resources for quantum simulator software, hardware, and education.

\textit{This document is a summary from a U.S. National Science Foundation supported workshop held on 16-17 September 2019 in Alexandria, VA.  Attendees were charged to identify the scientific and community needs, opportunities, and significant challenges for quantum simulators over the next 2-5 years.}
\end{abstract}

\maketitle

\section{Executive Summary}

Recent technical advances have brought us closer to realizing practical {\it quantum simulators}: engineered quantum many-particle systems that can controllably simulate complex  quantum phenomena. Quantum simulators can address questions across many domains of physics and scales of nature, from the behavior of solid-state materials and devices, to chemical and biochemical reaction dynamics, to the extreme conditions of particle physics and cosmology that cannot otherwise be readily probed in terrestrial laboratories~\cite{RevModPhys.86.153}. These state-of-the-art experiments involve the control of up to millions of  quantum elements and are implemented on a wide variety of atomic, molecular, optical, and solid-state platforms. Each architecture is characterized by strengths and weaknesses for solving particular classes of quantum problems. Simulators run the gamut from special purpose to highly programmable devices. These systems have the potential to fill a critical gap between conventional supercomputers---which cannot efficiently simulate many-particle quantum systems---and fault-tolerant scalable digital quantum computers, which may be decades away. At the same time, there are significant opportunities for co-development of the science and technology underlying both quantum simulators and fault-tolerant quantum computers.

Many years of progress by single investigators have evolved the field to a tipping point where investing in programs that bring together experimental scientists, theorists, computer scientists, and engineers will yield increasingly high returns and transformational outcomes. National leadership in quantum information science and engineering can be significantly bolstered by advancing this technology toward platforms that tackle pressing fundamental  and applied problems. We envision the creation of a robust ecosystem in this area that spans all sectors. Academic researchers will excel at stimulating technical and scientific breakthroughs, industry will lead transitioning research to commercial-scale systems for widespread availability, while national laboratories will contribute to shaping a large-scale R\&D effort, across the boundaries of traditional disciplines.
Industry, universities, and national labs will collaborate to focus on problems aligned with the long-term interests of end-users and society. Successful realization of this unique opportunity requires a dedicated national quantum simulator program.

We believe that realizing the ultimate potential of quantum simulators requires creating such a national program centered around \textbf{two main pillars}.
(1) \textbf{Early Prototype Quantum Simulators} will support the development, realization and deployment of complementary quantum simulator prototypes. They will  leverage --- rather  than  duplicate --- the  substantial  industrial  investment  in  technologies  and  software for digital quantum computing toward realistic, near-term simulator machines. This strategy will promote and establish access to the most mature architectures, and the software to operate them, by the broader scientific community to foster and accelerate practical quantum simulations.  (2) \textbf{New, Emerging Quantum Simulators} will support creative, cutting-edge research in science and engineering to uncover new paradigms, advance nascent hardware platforms, and develop new algorithms and applications for a new generation of quantum simulators. This effort will further support the development of new materials and devices to help accelerate the progress of new technologies and push them outside the research laboratory.
Here we document the opportunities and challenges in quantum simulators and explain our vision for accelerating the evolution of and capitalizing on this promising quantum technology via this two-pillar approach.

\section{Introduction}
\label{sec:introduction}

Quantum simulators strive to solve scientific problems that are not tractable by other means.  For the purposes of this article, we define a quantum simulator as ``\textit{a quantum device that utilizes entanglement and other many-particle quantum phenomena to explore and solve hard scientific, engineering, and computational problems}.'' Quantum simulators can be realized via different approaches, including highly tunable ``analog'' systems that naturally realize the physics problem of interest, or more digital methods that employ external control fields to produce non-native Hamiltonian evolution. Recent advances in several physical architectures promise a golden age ranging from highly optimized special-purpose quantum simulators to flexible and fully programmable devices.  These developments are enabled by a convergence of ideas drawn from fundamental physics, computer science, and device engineering.

Starting around 2002~\cite{greiner2002quantum}, early implementations of ultracold-atom-gas based quantum simulators  began to shed light on a range of long-standing problems such as quantum phase transitions, strongly interacting bosons and fermions~\cite{ingu08varenna}, the quark-gluon plasma, neutron stars, and the only measured quantitative new predictions from string theory (via holographic duality)~\cite{Adams_2012}. Quantum simulators have led to discoveries of unanticipated dynamical quantum many-body phenomena not governed by thermodynamic laws~\cite{schreiber2015observation}, and even realized systems without natural analogs such as hyperbolic space-times~\cite{kollar2019hyperbolic} and synthetic dimensions~\cite{lin2011spin}. Since many of these discoveries extend beyond what can be efficiently simulated exactly on a classical computer, quantum simulators are believed to exhibit a form of ``quantum advantage'' for computation.

Bulk-gas neutral atom experiments involving millions of particles~\cite{RevModPhys.80.885,Adams_2012}, enabling precision thermodynamic measurements e.g. on strongly interacting fermionic matter~\cite{Ku2012FermiEOS,Houcke2012}, are complemented by more bottom-up approaches with fully resolved control over smaller numbers of elements. Highly coherent quantum systems consisting of 50 or more individually addressed quantum elements have been created with superconducting circuits, trapped ions, and atom arrays. Recent improvements in materials, engineering, quantum control, measurement, and environmental isolation are opening new directions of study in these systems, including far-from-equilibrium many-body physics and large-scale entanglement. For example, novel ordering of spins was observed in a collection of over 50 Rydberg atoms held in optical tweezers~\cite{bernien2017probing}, a many-body order parameter was measured with over 50 trapped atomic ions to show the behavior of a dynamical phase transition~\cite{Zhang2017Nature}, and there has been impressive progress toward demonstrating quantum advantage and supremacy in superconducting circuits~\cite{arute2019quantum}.
Quantum simulations with ultracold fermionic atoms in optical lattices simulate condensed matter Hubbard models, and are shedding light onto pseudogap~\cite{chiu2019string} and strange metal physics~\cite{Nichols2019Spintransport,Brown2019Badmetal}. Meanwhile, novel quantum simulator architectures---e.g., gated quantum dots and photonic arrays---are on the cusp of opening unique and important new areas of investigation, such as modeling many-body quantum transport in engineered reservoirs and predicting properties of quantum complex networks for a quantum internet~\cite{vuckovic-qs,englund19}.

Progress on advancing quantum simulation techniques will be heralded by finer control over bulk systems and scaling addressable systems to work with larger number of particles. Both approaches have already established a frontier: understanding quantum simulators even at the 40--50 qubit scale is beyond direct brute-force diagonalization methods on classical computers. Scaling existing bottom-up quantum simulators to hundreds or even thousands of interacting, entangled, and well-controlled quantum elements \textbf{is realistically within reach.}

Achieving more advanced quantum simulation techniques will impact a wide range of fundamental and applied science and engineering. Efforts to create new materials by design will benefit from validated models of strongly correlated solids. New materials will offer potentially transformative properties, such as enhanced high-temperature superconductivity,  topological excitations,  unprecedented functionality for energy transport and storage, more efficient transistors, and advanced sensing technologies, as well as fundamentally new materials possessing properties with yet-to-be-conceived applications. Better understanding of non-equilibrium phenomena could lead to methods for generating light-induced topology and superconductivity, new techniques for control over chemical reaction kinetics, novel approaches to build robust quantum sensors for fundamental physics \cite{Campbell2017} and to enhance non-linear optical response, a more complete theory of the quantum origin of thermodynamics, and new insights into quantum gravity and lattice gauge theory. Applications to computer science could include hybrid digital/analog quantum computing, quantum approaches to combinatorial optimization problems, and quantum machine learning \cite{Biamonte2017}. Realizing this promise requires convergence of multi-disciplinary experimental and theoretical work, including engineering expertise throughout the design, implementation, and process deployment.

Fundamental research by single investigator groups continues to generate important advances in quantum simulation. Given the diversity of possibilities, and unpredictable nature of how quantum simulation will advance, it is essential to foster such research. However, quickly transitioning from fundamental research to mature developments ready for widespread availability, while maintaining alignment with the long-term interest of end-users, requires support and coordination best fueled by the creation of an interdisciplinary, multi-organization, and multi-investigator convergence accelerator program.  \textbf{Such a program will break down barriers among  disciplines, achieve a cross-pollination of ideas, and provide a coordination of effort that is essential for making rapid progress in the realization of practical simulators, especially those usable by the broader community outside of single-investigator groups.}

We envision a convergence accelerator program supported by two main pillars:

\begin{itemize}

\item Pillar 1, \textbf{Early Prototype Quantum Simulators}, focuses on assisting the rapid development of practical (i.e., achievable within 2--5 years) quantum simulator prototypes that can be made broadly available to the scientific community. This effort will employ a systems engineering approach to advance the most mature technologies. Existing facilities, commercially available quantum systems, and open-source software will be used to accelerate the timeline to deployment.  Key goals include making systems with improved flexibility for users and developing algorithms to speed the initial practical applications of these quantum machines. Success in this pillar requires focused, interdisciplinary collaborations including experimental and theoretical physics, engineering, and computer science. Early engagement of industry and national laboratories is critical to provide and manage access to existing systems for the broader scientific community. These stakeholders can also assist other technologies close to deployment readiness to become publicly available.

\item Pillar 2, \textbf{New, Emerging Quantum Simulators}, focuses on accelerated convergence of fundamental science and engineering to discover principles, devices, and new applications that will lead to the next generation of simulators with improved capabilities. Priorities include theoretical work in computer science and quantum many-body physics, the development of new architectures and algorithms for quantum simulators, novel classical methods to simulate many-body systems informed by quantum simulator results, and continued fundamental research that advances existing and emerging quantum simulator platforms. This activity will require further work on experimental systems,  classical engineering  for electronic and optical control systems, developing new materials and device fabrication techniques to improve coherence, and work toward improving the programmability and scalability of quantum simulator platforms.
\end{itemize}

The remainder of this paper is organized as follows. Sec.~\ref{sec:fundamental} details the opportunities and challenges for quantum simulators to explore a range of key scientific problems.  Sec.~\ref{sec:architectures} overviews existing quantum simulator platform architectures. Sec.~\ref{sec:programmability} addresses  critical requirements to produce reliable simulators, particularly tunability, programmability, and verifiability. Sec.~\ref{sec:collaboration} describes the community structures that can take quantum simulators to the next level, including the formation of a quantum hub internet resource for quantum simulator hardware, software, and education.  Conclusions are provided in Sec.~\ref{sec:conclusions}.

\section{Using Quantum Simulators to Investigate Fundamental Science and Beyond}
\label{sec:fundamental}

Quantum simulators have the potential to solve a wide range of important outstanding scientific problems and dramatically impact technology development. They can tackle important models of quantum materials that defy understanding despite decades of intensive research, promise breakthroughs in accurate simulation of chemical processes far out of reach of even the most advanced supercomputers, and can access extreme conditions relevant to high-energy particle physics and cosmology otherwise impossible to create in the laboratory. Moreover, quantum simulators enable an unprecedented degree of spatio-temporal control and measurement detail and precision~\cite{Marti2018}, offering deeper insights than previously accessible.

Near-term advantages will necessarily be realized with systems that lack full error correction. The impact of imperfections such as noise and decoherence cannot be accurately modeled theoretically, and must be explored empirically by building and testing successively more complex simulation platforms. Constant experimental advances will be required to reduce noise, increase the number of interacting quantum elements, and improve flexibility and programmability. Theoretical and algorithmic advances are required to extend the capabilities of quantum simulators, identify new directions for
research, analyze and interpret experimental results, and synthesize outcomes from different platforms into a unified perspective. These challenges, detailed further below, highlight the need for cohesive collaborations between experiment and theory.

\subsection{Scientific opportunities for quantum simulators}

\textbf{Quantum materials simulation}:
Many deceptively simple models of correlated electronic materials defy efficient simulation due to sign problems associated with fermionic statistics, spin frustration, or complex gauge fields. A paradigmatic example is the fermion Hubbard model~\cite{lewenstein2007ultracold}, which is commonly regarded as a minimal microscopic starting point for describing the physics of a host of solid-state materials, from high-temperature superconductors to frustrated quantum magnets. Despite  decades of theoretical and experimental efforts, the physics of the Hubbard model and the materials it models are not fully understood. Spin and Bose-Hubbard systems have similar complexity in the presence of strong interactions and gauge fields or spin frustration. Quantum simulators can implement these models and explore exotic low-temperature phases and higher temperature regimes, including the famous pseudo-gap, strange metals, and the quantum critical fan.   Simulations of more complex features including heterostructures, artificial lattice structures such as quantum spin ice, and quantum generalizations of soft matter such as the spin glass are also achievable. Furthermore, quantum simulators allow more complete access to quantum states and quantities difficult to access in conventional materials, including non-local observables, entanglement measures, high-order correlators, and even full snapshots of the quantum wave-functions.  On the 2--5 year time scale a wide variety of such materials simulations can be performed in a quantum simulation version of the materials genome initiative, particularly as prototypes are made available to the materials science community via pillar 1, and as multi-investigator blended physics--materials science teams are formed in pillar 2.

\textbf{Quantum Chemistry:} Chemistry problems such as calculating reaction rates and modeling catalysis are tantalizing topics with societal importance, including more efficient nitrogen fixation~\cite{reiher2017elucidating} and making synthetic versions of light-harvesting photosynthetic complexes.  Classical computation proves insufficient in many areas, including reactions occurring through conical intersections, determining whether the reaction products are controllable by wavefunction manipulation of the reactants, electron transport and energy harvesting in multi-scale molecules and environments, and even many problems in bond energies and bulk properties in modest-sized molecules. For example, accurately calculating molecular properties of a single Cr$_2$ dimer from first principles remains beyond the reach of current techniques.  Quantum simulators can solve such problems by directly emulating a model of such molecules or reactions and are not limited by the high overhead required by digital quantum computers.  For example, recent progress has been made on simulating the interplay of molecular vibrations and electronic properties, and quantum simulations modelling phonon-assisted energy transfer have been carried out~\cite{gorman2018engineering}. Extending these simulations will enable the  identification of previously undiscovered electron--phonon interactions.  On a 2--5 year time scale a direct model of the photosynthesis problem can be built on several quantum simulator architectures to help resolve the ongoing controversy over the role of quantum coherence in exciton transport and energy harvesting~\cite{streltsov2017colloquium,ball2018photosynthesis}, with applications to a new generation of photovoltaics and many other chemical problems such as multi-scale enzyme reactions~\cite{meisner2016atom}.  The role of accessible prototypes to the chemistry community under pillar 1 will greatly enhance such studies.

\textbf{Quantum devices and transport}:  Similar to constructing a direct model of quantum processes in complex molecules, quantum simulators admit tunable models of quantum devices. Quantum simulators can play a key role in understanding the transport of current, spin, heat, and information and the influence of coupling to gates and substrates in quantum devices. A general understanding of transport needs to build beyond semi-classical limits and linear response to include quantized interactions and correlations.  For example, a complete quantum understanding of the basic building block of CPUs---the field-effect transistor (FET)---may provide both fundamental insight and identify physical limits; quantum simulators are already exploring this area~\cite{krinner2017two}. Recent experiments are starting to shed light on the transport properties of the Fermi-Hubbard model, believed to hold the key for the understanding of high-temperature cuprate phenomenology~\cite{Nichols2019Spintransport,Brown2019Badmetal}. Likewise, transport is essential to quantum-dot based photovoltaics, quantum thermoelectrics, spintronics, and other quantum entanglement-enhanced devices that tolerate or even prefer decoherence, rather than avoid it. The question of quantum thermoelectric effects and the quantum meaning of heat transport ties in to nanothermodynamics more generally~\cite{giazotto2006opportunities}. In this direction, creation of information-based engines have been proposed, with initial demonstrations of an information-based Carnot cycle~\cite{PhysRevLett.122.150604}, but a great deal of exploration is needed to unlock this field. Finally quantum complex networks can be realized in very recent quantum simulator architectures, allowing one to predict and explore properties of future ``quantum internet'' channels and device elements. On the 2--5 year time scale, environment-enhanced quantum transport in a wide variety of open quantum systems, quantum complex networks, and quantum nanothermodynamic devices can all be realized on several quantum simulator architectures. Progress in this area will especially benefit from multidisciplinary teams combining electrical engineers and physicists.

\textbf{Gravity, particle physics, and cosmology}: Quantum simulators will serve as laboratory test-beds for non-terrestrial phenomena relevant to particle physics, cosmology, and quantum gravity. Simulators can explore a wide range of phenomena, including lattice gauge theories~\cite{martinez2016real}, color superconductivity, cosmological defect production in inflating spacetimes and quantum effects in curved spacetimes. Investigations into the dynamics of thermalization have revealed striking and unanticipated connections linking aspects of many-body quantum chaos, scrambling of quantum information theory, holographic descriptions of black-holes, and quantum gravity~\cite{swingle2018unscrambling}. These conjectured links can be directly tested in quantum simulations, forging new connections between these seemingly disparate disciplines. Entangled quantum many-body systems constructed for simulation purposes, combined with high-precision measurements, could even act as highly sensitive detectors to explore gravitational effects.  On the 2--5 year time scale, the question of information scrambling in out-of-time-order correlators can be addressed on near-term quantum simulator architectures, particularly with the added benefit of multi-disciplinary supported collaboration between cosmologists and string theorists together with quantum simulator theorists and experimentalists.  A national quantum simulator program can also support joint programs in this area (see Sec.~\ref{sec:collaboration}).

\textbf{Non-equilibrium quantum many-body dynamics}: Quantum simulators also enable access to fundamentally new regimes of coherent, closed-system, and highly excited dynamics previously inaccessible in conventional materials~\cite{eisert2015quantum,abanin2019colloquium}. This important problem spans all scales of nature, from the formation of large-scale structures in the universe to the response of semiconductor devices. The potential impact of quantum simulators in this domain is high, since dynamics and excited states for multi-particle interacting quantum systems are largely inaccessible to conventional computing. Furthermore, theoretical tools for describing out-of-equilibrium behavior are limited. Quantum simulators can access the full range of many-body dynamics, including relaxation to thermal equilibrium in some closed systems, to non-thermalizing behavior in integrable and disordered systems~\cite{abanin2019colloquium}. Open challenges addressable by quantum simulators include uncovering the universal organizing principles governing out-of- and non-equilibrium quantum phenomena;  understanding how time-dependent control, driving, and engineered dissipation can be used to stabilize non-equilibrium quantum phases~\cite{harper2019topology}; and enabling error-resilient storage and manipulation of quantum information~\cite{omran2019generation}. In thermalizing systems, quantum simulators can shed light on transport properties and non-linear response, which are often poorly understood despite their critical importance for characterizing materials and designing electronic devices.  Understanding non-equilibrium quantum many-body dynamics is perhaps the greatest strength of quantum simulators and ties together all the topics of this section, falling squarely into the area of pillar 2 under fundamental research.

\subsection{Challenges and Opportunities for Theory}

Theoretical work plays a critical role in designing quantum simulators, interpreting the experimental data they produce, and developing an overarching framework to apply lessons learned in synthetic simulators to physical systems of interest. The new theoretical ideas and methods needed to drive breakthroughs in quantum simulation are possible only through a symbiotic relationship between theoretical and experimental teams.

Advances in quantum simulators pose both challenges and opportunities for theory. One key challenge is the need to move beyond existing frameworks by investigating new regimes of non-equilibrium dynamics not described by standard statistical mechanics and capitalizing on access to non-local observables such as  entanglement~\cite{islam2015measuring}, string orders, snapshots of many-body wave functions~\cite{bakr2009quantum}, and distribution functions. Standard theoretical approaches to many-body systems emphasizing two-point correlation functions of local observables and near-equilibrium linear response must be extended. Needed advances include effective field theories, variational methods, and advanced numerical techniques.

An opportunity for theory is the possibility for detailed comparison of different model predictions against quantum simulation experiments, which
will be fundamental to extract lessons about the quantum problems at stake. For example, in the case of quantum materials, many classes of theories have been proposed to explain competing orders, quantum criticality, and proximity to exotic fractionalized phases in high-temperature superconductors. These theories can be used as a foundation for model-based analysis of the complex data sets from quantum simulation experiments. Such a program requires new methods to translate the predictions from these models into the observables from quantum simulators. Advances in data-science and machine-learning tools can also be used to compare theory and experiment and contrast different approximate theories. The same needs apply to other fields, such as interpreting nuclear magnetic resonance (NMR) spectra in biomedical systems.  Successes on quantum simulator problems can thus translate into understanding in other fields, such as NMR. Success calls for organizing multi-disciplinary collaborations between experts in data science, theoretical physics, medicine, engineering, and other fields of science.

\subsection{Challenges to building a large-scale quantum simulator}

The main technological challenges to leverage the strengths and overcoming the weaknesses of quantum simulators are common across platforms and center around scalability, complexity, state preparation, control, and measurement.

\textbf{Scalability and complexity}:
The quantum simulator architectures, numbered A--1 through A--11, are described in detail in Sec.~\ref{sec:architectures}.  Some platforms, such as A--2 (color centers), A--3 (dopants in semi-conductors), A--4 (gate-defined quantum dots), and A--8 (superconducting quantum circuits), face the challenge of variability, i.e., quantum devices with nominally identical design rarely exhibit identical performance, are affected by coupling to nearby environments or defects, and typically require time-intensive independent calibration. Addressing this issue will require better understanding of imperfections and coupling to the environment; developing better techniques for placement and control of impurities, donor atoms, and vacancy centers that act as artificial atoms; and devising automated calibration processes to tune-up a large number of independent devices. Furthermore, many problems of interest require long-range connectivity that typically cannot be easily be achieved by physical wiring in planar devices. Implementing long-range interactions via SWAP networks is technically possible, but often incurs a large overhead. Progress is needed to enhance physical connectivity of devices, and devise better algorithmic methods to improve connectivity. Along with connectivity comes cross talk or unwanted connections and interactions, as one finds natively in many of the other architectures, e.g. in A--1 (cold and ultracold molecules), A--6 (photons and atoms in cavities), A--7 (Rydberg atom arrays), and A--9 (trapped atomic ions). A careful design of the system architecture is required so that cross-talk errors do not scale with the system size. As discussed below, as the system size grows, so does the complexity, not only for the physical architecture and the control and measurement hardware, but also for the quantum states involved and their verification.

\textbf{State preparation and control:}
As quantum simulators enter ever larger scales and complexities, it becomes increasingly difficult to reach and maintain the low entropy and low effective temperatures required to prepare and investigate strongly correlated quantum many-body states. Multiple platforms will benefit from new methods of state preparation, including new cooling techniques, entropy redistribution via reservoir engineering, stabilization via a dissipative bath, and measurement-based state preparation.
Alternatives to traditional adiabatic protocols include (i) autonomous error mitigation for analog quantum simulators using engineered baths/dissipation e.g. in architectures A--7 (Rydberg atom arrays), A--8 (superconducting quantum circuits), and A--10 (ultracold neutral atoms); (ii) new tools for  quantum state control, see Sec.~\ref{sec:programmability};
and (iii) optimizing state-preparation protocols, cooling, and variational approaches in near-term algorithms. These challenges present unique opportunities for disparate disciplines such as computer science, electrical engineering, materials science, and mathematics to contribute critical insights making large scale quantum simulation possible.

\section{Quantum Simulator Architectures}
\label{sec:architectures}

Quantum simulators can be assembled from a wide variety of atomic, molecular, optical, and solid-state physical platforms.  In the following, we summarize these platforms in alphabetical order.

\vspace{.1in}
\textbf{A--1: Cold and Ultracold Molecules}
\vspace{.1in}

Polar molecules present a  quantum element combining the strong, anisotropic electric dipolar interaction with one to hundreds of internal states with transitions at convenient frequencies~\cite{carr2009cold,gadway:strongly_2016,koch:quantum_2019}.  They can be assembled from neutral atoms to reach Fermi degeneracy~\cite{de2019degenerate}, or be directly laser cooled~\cite{doi:10.1080/00107514.2018.1576338}. Molecule-based quantum technology will improve precision measurement~\cite{RevModPhys.90.025008} and provide insights in molecular physics, chemistry, and biology. Their range of frequencies, kHz to hundreds of THz, lends itself to efficient information transduction across different platforms and study of multiscale quantum physics, while combining with their intrinsic tunable interactions will allow them to reach high-fidelity entanglement. Stacked multi-layer fermionic polar molecules~\cite{Miranda2011} can simulate the extended-Hubbard and $t$-$J$ models and  topological phases, while different classes of molecules can explore models difficult to access with other platforms, e.g. exotic XYZ magnetism, and internal states can provide a separately tunable reservoir for simulations of open quantum systems.  Crucial challenges for lattice-based quantum simulators include cooling the system to the many-body ground state, obtaining more than about $10^3$ to $10^4$ quantum elements, and extending the classes of molecules to simulate more models.  Addressability at the single molecule level is another key goal, as well as trapping and control of cold molecules in optical tweezers~\cite{liu2019molecular,anderegg2019optical}.

\vspace{.1in}
\textbf{A--2: Color Centers}
\vspace{.1in}

Color centers in wide band gap semiconductors (diamond, SiC, etc.) feature microwave transitions for control and optical transitions for coupling to each other and to cavities~\cite{diamond-sic-review}.
This system does not require high vacuum or particle trapping, and performs well at modest temperatures ($>1.5\ {\rm K}$)  with coherence times ranging from hundreds of microseconds to minutes. Small quantum registers of up to 10 fully-connected qubits have been recently demonstrated~\cite{bradley2019ten}.
Color centers are well suited for the study of open quantum systems~\cite{lei2017decoherence}. Two quantum registers have been optically connected using nitrogen-vacancy centers in diamond~\cite{humphreys2018deterministic}. Active areas of research to facilitate scalability include increasing the quality of optical interconnects through improved fabrication methods and photonic optimization~\cite{dory2019inverse,dlukin-sic},
exploration of new defects, integration of defects to photonics structures, and precise control over defect placement.

\vspace{.1in}
\textbf{A--3: Dopants in Semiconductors}
\vspace{.1in}

Dopant atoms have been integrated into gate-defined quantum-dot devices using deterministic ion implantation~\cite{RN1339}, and have been placed with atomic precision into dopant-based devices using scanning probe fabrication. Both single-qubit and two-qubit gates have been shown using the electron spin bound to donor atoms~\cite{He2019Two}. Emergent many-body behavior has been seen in the transport through short 1D chains of donors~\cite{Prati2016Band}. In general, dopants in semiconductors have many of the same opportunities and limitations of gate-defined quantum dots, with some notable differences. The nuclear spin provides an additional degree of freedom that has very long coherence times, and can be coupled to bound electrons~\cite{HarveyCollard2017Coherent}. The main challenge of this platform is achieving tunable donor-donor coupling. Scaling to larger arrays also requires managing the wiring for individual qubit control.

\vspace{.1in}
\textbf{A--4: Gate-defined Quantum Dots}
\vspace{.1in}

Spin qubits in gate-defined quantum dots offer a tunable, scalable platform with long spin coherence relative to their gate time~\cite{vandersypen2019quantum}. Advantages of this technology include the ability to scale to moderate numbers of qubits using nanoscale lithographic patterning on high-quality semiconductor silicon,  and silicon-germanium substrates.
1D and 2D arrays of these structures have been demonstrated with excellent control.  Silicon quantum dot platforms have achieved single-qubit and two-qubit gate fidelities of 99.9\%~\cite{yoneda2018quantum} and 98\%~\cite{huang2019fidelity} respectively. Strong coupling between double quantum dot qubits and superconducting coplanar waveguide resonators has recently been achieved.
A Fermi-Hubbard model quantum simulator has been implemented with quantum dots~\cite{hensgens2017quantum}.
Quantum dots offer the possibility of long-range interactions with controllable disorder for simulation of quantum spin fluid formation, Wigner crystallization, Anderson localization, and time crystals, as well as opportunities for reservoir engineering for open quantum systems.
Active areas of research include exploring material science parameters to reduce inter-device variability, methods of using variability as a resource, modified measurement schemes to improve speed and scalability, such as multiplexed dispersive readout, and automation of tuning and readout protocols.

\vspace{.1in}
\textbf{A--5: Photons in Nanostructures}
\vspace{.1in}

Engineered nanophotonic structures coupled to strong nonlinearities
offer a potential solution to realizing tunable long-range interactions in two-dimensional lattices~\cite{gonzalez2015subwavelength}. This will facilitate simulation of quantum spin fluid formation or Wigner crystallization, and Bose Hubbard models~\cite{hartmann2008quantum}. Furthermore, the spectral or spatial inhomogeneity of the quantum emitters providing the nonlinearity also acts as a disorder term, making these platforms well suited for implementation of disordered many-body systems with long range interactions,
challenging to investigate classically since the analytical tools used to study symmetric and translationally invariant systems break down in the presence of disorder. Optimization techniques will play an important role in engineering of photonic structures~\cite{molesky2018inverse}.

\vspace{.1in}
\textbf{A--6: Photons and Atoms in Cavities}
\vspace{.1in}

Cavity quantum electrodynamics (QED) enables coupling of distant atoms through photon exchange via the mode of a cavity operating in the strong-coupling regime.
These long-range interactions, which can include all-to-all connectivity, allow simulation of exotic spin models \cite{Vaidya2018,Norcia2018,Davis2019}.  Coupling to several cavity modes and/or Floquet driving provide adjustable range, connectivity, and sign of the interactions, which in turn enable simulation of phenomena ranging from quantum spin glasses to information scrambling in quantum black holes. Alternatively, these setups can be viewed as many-body systems of cavity photons with strong photon-photon interaction mediated by atoms~\cite{Noh_2016, Hartmann_2016}, and Laughlin states of light have been recently demonstrated~\cite{Clark2019Laughlin}. In current experiments, the cavity mode often couples collectively to large clusters of atoms, which therefore behave as large semi-classical spins with cooperatively enhanced interactions. This regime has allowed for deterministically generating many-particle entangled states, including squeezed \cite{leroux2010implementation,hosten2016quantum,braverman2019near,PezzeQMetrology} and $W$ states \cite{barontini2015deterministic}.  In an opposite limit, cavity-mediated interactions have enabled a two-qubit entangling gate \cite{welte2018photon}.  An important challenge for future experiments is to go deeper into the  quantum regime in scalable many-body systems, by combining strong atom-light coupling with local tunability of the interaction of individual atoms with the cavity.

\vspace{.1in}
\textbf{A--7: Rydberg Atom Arrays}
\vspace{.1in}

Deterministically prepared, often reconfigurable 1D, 2D and 3D arrays of individually trapped cold neutral atoms with strong, long range, coherent interactions enabled by excitation to Rydberg states constitute a promising approach to realizing programmable quantum simulators.  Recent experiments realized quantum spin models using over 50 qubits with tunable interactions~\cite{bernien2017probing}, leading to the discovery of a new class of quantum many-body states that challenge traditional understanding of thermalization in isolated quantum systems. The work also provided insights into  quantum phase transitions in exotic, previously unexplored condensed matter models. Symmetry-protected topological phases of matter have been realized using 2D atom arrays \cite{Leseleuc2019}.
High-fidelity single-qubit rotations and multi-particle entangled states, quantum state detection, and parallel quantum logic operations have been demonstrated, establishing neutral atoms as a competitive platform for quantum information processing.
Challenges include scaling to hundreds of entangled quantum elements and reproducing results on similar or complementary platforms.

\vspace{.1in}
\textbf{A--8: Superconducting Quantum Circuits}
\vspace{.1in}

Superconducting circuits are a leading solid-state quantum simulation platform
with relatively long coherence times and strong, tunable interactions enabling fast, high-fidelity quantum logic gates. Relatively high-fidelity operation and readout of more than $50$ qubits is now achievable~\cite{arute2019quantum,IBM53qubit}.
A wide range of circuit topologies can be lithographically defined, making superconducting circuits one of the most geometrically flexible simulator platforms~\cite{NoriReport2017}. Quantum elements include linear resonant elements operating as cavities or photon (bosonic) reservoirs~\cite{michael2016bosoncode}, and  nonlinear elements operating as artificial multi-level atoms (qudits), two-level  “atoms”, “spins”, or qubits~\cite{NoriReport2017,kjaergard2019scc}. Linear and nonlinear couplings between elements can provide tailor-made interactions~\cite{NoriReport2017,krantz2019guide,MGanzhorn2019} for investigating frustrated spin systems, photon-photon interactions, or modelling phase-transitions beyond mean-field theory. This platform has been applied to digital simulation of spin models, fermionic dynamics, quantum chemistry using variational quantum eigensolvers; and machine learning for classification~\cite{GWendinReview2017,MGanzhorn2019,vqe_small_molecules,classifier,google_vqe}. Analog simulation of strongly correlated quantum matter --- such as many-body localized states in disordered lattices and the Mott insulating phase in the Bose-Hubbard model~\cite{Ma2019dissipatively} --- has been recently explored. Significant challenges in this platform include obtaining high fidelity and full quantum control over more than tens of qubits, while maintaining and improving the coherence of individual circuit elements.

\vspace{.1in}
\textbf{A--9: Trapped Atomic Ions}
\vspace{.1in}

Trapped atomic ions are among the most coherent and controllable of all quantum simulation platforms~\cite{blatt2012quantum}. The internal structure (spin) of atomic ions is not perturbed by the external confinement forces.  Trapped ion qubits can be localized, initialized, and measured with high fidelity. Reconfigurable qubit couplings are mediated by the Coulomb interaction and controlled through state-dependent dipole forces realized with applied optical or microwave fields. Qubits can be fully connected for modest system sizes.  This allows the implementation of magnetic spin models with long-range and tunable interactions, and simulations of phase transitions and quantum dynamical processes through adiabatic manipulations, quantum quenches, and Hamiltonian modulation~\cite{hess2017non}. Variational quantum eigensolver algorithms for quantum chemistry have also been demonstrated with this platform~\cite{hempel2018quantum, nam2019arXiv}.
Current state-of-the-art trapped ion quantum simulators involve more than 50 spins \cite{Britton2012Nature,bohnet2016quantum,Zhang2017Nature}, with complete control and high fidelity up to a few tens of ions. The primary challenge to scaling ion traps is the systems-level engineering of optical controllers and electrode structures to connect groups of trapped ions through shuttling, as well as the integration of photonic interconnects for modular communication between trapped ion qubit clusters.””

\vspace{.1in}
\textbf{A--10: Ultracold Neutral Atoms}
\vspace{.1in}

Ultracold neutral atoms working with thousands to tens of millions of interacting fermionic or bosonic quantum elements can simulate a wide variety of paradigmatic quantum phases and phenomena because of their high degree of isolation, flexible geometry, interaction and spin control, and slow observable dynamics. Fermionic degrees of freedom can be natively realized. Quantum simulator tools range from optical lattices~\cite{jaksch2005cold,lewenstein2007ultracold} and Feshbach resonances realizing strong interactions~\cite{ingu08varenna,Chin2010Feshbach} to now artificial gauge fields, arbitrary confining geometries, dipolar interactions, coupling to optical cavities, and individual atom imaging and control in quantum gas microscopes~\cite{bakr2009quantum,Cheuk2015Fermi}, allowing one to directly sample the many-body wave function.
Equilibrium quantum simulations accessible to these devices include the Fermi-Hubbard model (reducing temperature roughly four-fold may exhibit the predicted $d$-wave superconductivity), quantum spin liquids (using lattices with geometrical frustration),  fractional quantum Hall physics (using artificial gauge fields), and even models from cosmology (using dynamical geometry or interaction strengths).  Nonequilibrium simulations include the breakdown of thermalization in isolated integrable systems, the dynamics of many-body localization, and high order correlators, for example in atom interferometry~\cite{langen2015experimental}. The unified approach of quantum simulation and precision metrology~\cite{Martin2012,Kolkowitz2017} is leading a revolution in our understanding of complex systems for applications in measurement science. Floquet engineering on atomic quantum gases has already synthesized gauge fields relevant to condensed matter, high-energy, particle and gravitational physics.  Challenges include the non-uniformity inherent to many trapping scenarios, a variety of loss and decoherence mechanisms from background vacuum gas to trapping fields, and removing entropy to achieve ground states.

\vspace{.1in}
\textbf{A--11: Van der Waals Heterostructures, Moir\'e Materials, and Excitons}
\vspace{.1in}

Atomically thin 2D materials such as graphene and transition metal dichalcogenides (TMDs) can be combined into layered heterostructures of great variety. These structures effectively simulate a highly tunable Hubbard model, analogous to what occurs in complicated correlated materials, but on a greatly expanded length scale. The relative importance of electronic tunneling and correlations can be continuously varied by tuning gate voltages, relative interlayer twist angles, strain, and dielectric environments, to explore a full phase diagram in a single material~\cite{macdonald2019trend}.
Moir\'e superlattice structures of graphene have realized highly tunable correlated insulators, magnets, and superconductors, and future heterostructures will be able to simulate even more complex correlated topology, frustrated magnetism, and fractionalized spin liquids.
2D material heterostructures can also host high densities of long-lived bosonic particles such as excitons and exciton-polaritons, manipulable
to produce highly tunable solid-state platforms to simulate bosonic many-particle systems, and to realize optoelectronic and thermoelectric applications~\cite{wang2018colloquium}.
These architectures can be cooled more easily, relative to the tunneling and superexchange energies, than for instance their counterparts in architectures A--1 (cold and ultracold molecules) and A--10 (ultracold neutral atoms) to explore coherent spin-physics, presenting both an exciting opportunity and a challenge.

\section{Programmability and Verifiability}
\label{sec:programmability}

Quantum simulators can take many forms ranging from completely analog to fully digital, with hybrid implementations in between~\cite{kokail2019self}. Having presented the scientific challenges and opportunities in Sec.~\ref{sec:fundamental}, and the architectures in Sec.~\ref{sec:architectures}, this section proceeds to identify opportunities and challenges in the programmability and operation verification~\cite{hauke2012can} of this large class of simulators.

\subsection{Digital, Analog, or Hybrid}

While quantum simulators are traditionally classified as analog or digital, actual realizations are more nuanced, spanning the continuum between the two.
Digital, i.e., gate-based quantum simulation, has the advantage of versatility since in principle, all Hamiltonians can be effectively encoded into quantum circuits involving just one- and two-qubit gates. However, the high coherence and gate fidelity, along with error correction, required for deep quantum circuits are unattainable in the next 2--5 years. Quantum simulators spanning the continuum between digital and analog provide a more realistic path to solving scientifically pressing problems in the near term.

Fully analog devices fix the Hamiltonian to mimic the problem exactly. Adding tunability allows broadening to a class of problems rather than a fixed Hamiltonian. Additional capabilities such as single-qubit gates or single-site addressability provide added control and an even larger class of problems to be tackled. On the other hand, in a model closer to the digital quantum computer, allowing for analog-like elements such as switching on/off multi-qubit interactions, rather than decomposing multi-qubit gates in terms of two-qubit and single-qubit gates, can lead to shorter depth circuits and the ability to simulate more complex problems, despite limited coherence and gate fidelity.

Hybrid models of simulation can also open new opportunities. This can include either hybrid approaches involving different architectures from Sec.~\ref{sec:architectures}, or quantum-classical hybrid simulations, such as variational quantum eigensolvers. In the former case, one approach is running the same problem on different hardware to cross check and benchmark the results; another possibility is a physical connection between disparate systems via a quantum channel. For quantum-classical hybrid simulators, the quantum system is used for encoding and evolving the quantum state as an ansatz, while the classical computer is used for optimization of its parameters.

\subsection{Challenges and Opportunities}

Before problems of practical importance can be tackled, it is imperative to test the ability to accurately control engineered quantum systems, verify that the results of the quantum simulators can be trusted,
and quantify errors.
Gate-based quantum simulators benefit from verification and validation methods developed for quantum computing, such as randomized benchmarking and gate set tomography, to characterize system performance and diagnose errors. However, these tools have an intrinsic exponential scaling, limiting them to smaller systems. For analog quantum simulation, the theoretical design of benchmarking and error-correction tools is in its infancy. For example, what is the equivalent of a ``quantum volume''~\cite{bishop2017quantum,preskill2018quantum} for quantum simulators that will enable comparison of the different architectures? There is a broad opportunity for witness observables, akin to entanglement witnesses, to validate simulator operation without a priori predictions for the outcome of the complete simulation. In fact, we expect techniques for validating, calibrating, and diagnosing quantum simulators to be platform agnostic. Similarly, quantum error correction methods developed for universal quantum computers provide high-level protocols transcending specific physical architectures but require overheads too large for near-term simulators. Uncovering new types of error correction and mitigation suitable for near-term simulators is an open problem.

Furthermore, there is a natural trade-off between the coherence of a system and the degree of programmability and tunability. Understanding and exploring this trade-off is important in developing quantum simulators. There is also an opportunity in the development of algorithms with this trade-off in mind, leading naturally to the concept of co-design, i.e., developing applications for specific hardware and architectures. Such efforts will benefit from the convergence of various areas of expertise, including experimental physics, quantum control theory, computer science and software, and engineering.

\textbf{Validating Analog Simulators}:
Physical systems contain contributions to their Hamiltonian beyond the strict confines of the model that they are designed to implement~\cite{tomadin2007many}, representing a challenge to analog quantum simulators. However, exact agreement between a simulator and its target model is not expected to be necessary. Thanks to the universality of the low energy physics arising from collective behavior, often these additional terms or fluctuations will not affect the correctness of the simulation provided they are not too large. Current experiments show that good agreement with simple and tractable theoretical models can be achieved despite the presence of imperfections~\cite{bloch2012quantum}. An immediate question is whether it is possible to quantify the robustness of observables to such perturbations. Classification of the relevance of perturbations to the systems as relevant, marginally relevant, or irrelevant bear resemblance to classification schemes in renormalization group flows~\cite{beny2015information}. Methods to determine how strongly observables are affected by small perturbations and to derive reliable error bounds are desirable.
This presents remarkable research opportunities to quantify uncertainties in systems without a-priori known behavior.

\textbf{Validating Digital Simulators:} Digital quantum simulation suffers from several inherent sources of errors, including digitization errors arising from insufficient number of Trotterization steps, control errors on the gate unitaries, and decoherence due to both dephasing and environmental interactions.
Well-understood tools exist for characterizing and dissecting these errors in small-scale quantum circuits, such as randomized benchmarking, cross-entropy benchmarking, and state and process tomographies.  However, these tools  become forbiddingly difficult to implement at the level of 10 to 50 qubits, let alone for larger and more complex quantum elements \cite{eisert_quantum_2019}.
Possible directions will involve new benchmarking protocols to estimate the fidelity of quantum circuits without explicit reliance on classical calculations, and new measurement methodologies improving upon the exponential or super-exponential scaling behavior in the number of quantum elements in the system.
The use of computer-science validation techniques will also prove crucial.
For example, machine-learning based techniques can be used to validate and reconstruct properties of large quantum hardware \cite{carleo_machine_2019}. Success in any of these efforts, even moderate, will have profound impact on digital quantum simulation and computation as a whole.

\textbf{Comparison to Classical Calculations}:
Comparison of experimental results with controlled classical simulations of quantum systems plays an important role in the validation of quantum hardware.
For small systems, brute-force classical simulations can be still performed without essential approximations.
At this scale, systematic verification and comparison procedures can reveal flaws or neglected relevant perturbations in microscopic models.
For larger-scale quantum hardware, comparing to classical simulation is more challenging but venues for benchmarking are still open. State-of-the-art classical many-body techniques can be used
to validate analog and digital quantum computers. For example, low-entangled states in 1D are naturally amenable to tensor-network simulations \cite{schollwock_density-matrix_2011}, whereas sign-problem-free models can be effectively studied with stochastic Quantum Monte Carlo techniques \cite{foulkes_quantum_2001}. Comparing to classical simulations has a dual goal. First, it is crucial in understanding the microscopic details of quantum hardware and better assessing its capabilities. Second, experimentally accessing regimes beyond the current applicability of many-body classical techniques can be regarded as a way to benchmark and improve upon such techniques.

\textbf{Verification of Simulators}:
Given that many problems are well-suited to implementation on multiple platforms,
comparison of simulations of the same Hamiltonian using different technologies
can verify the outcome or reveal unexpected interactions in the simulators.
Alternatively, one can use self-verification, where the simulator is run
forward and then backward, testing that the state at the midpoint has the desired properties of the simulation and that the inversion of the evolution returns to the same initial state.  A more complex self-verification will generate the same Hamiltonian through two different physical processes to compare outcomes.

\textbf{Error correction and mitigation for quantum simulators}:
The type of quantum simulator hardware determines the impact and types of imperfections affecting the system. Decoherence and unwanted interactions have a different physical origin depending on the platform. The central challenge is to identify and correct for these unwanted interactions. To date, the most successful approach is to address every imperfection individually by careful engineering and ingenuity.
A methodology is needed for quantum simulators that, similar to quantum error correction, is able to mitigate the effect of errors and offer a high degree of portability across different architectures.
Because quantum simulators do not demand the same flexibility as universal quantum computers, they open up the possibility to devise error mitigation schemes that are targeted at special purpose simulations with near-term devices~\cite{temme2017error,li2017efficient,kandala2019error,mcclean2019decoding}. More research is needed to explore opportunities to apply methods derived from quantum error correction and mitigation to analog simulators.

\textbf{Mesoscopic Metrics of Quantum Complexity}: Most models of computational complexity are based on the asymptotic limit of computation, an approach that is relevant for large systems. This is a poor model for quantum simulators, which possess both limited control and limited components. Developing a complexity theory of quantum simulation that is more relevant to medium-scale systems is an open problem that will impact the further design of these platforms.

\section{Fostering Collaboration and Shared Resources}
\label{sec:collaboration}

Ultimately, the goals of pillar 1 (to create a transition of mature quantum simulator architectures from laboratory demonstrations to prototypes to end-user products accessible by the broader scientific community) and pillar 2 (to bring less well-developed platforms up to that point and perform fundamental research) require a multi-faceted program to support multi-disciplinary team formation beyond the present predominantly single investigator level.  In the following, we describe our vision of community support for such a national quantum simulator program.  It is the position of the community that this approach is necessary to succeed on a 2-5 year time scale.

\subsection{Horizontally and Vertically Integrated Teams}

The design and operation of a quantum simulator involve a high level of collaboration between theorists who introduce hard models, computer scientists and systems engineers who ensure that the simulator is appropriately tunable, reliable, and usable, and quantum experimentalists who implement and run the device and extract the data.  There are common techniques among multiple physical architectures, from manipulating individual photons, to trapping and interrogating atoms and molecules, to controlling decoherence in quantum circuits and materials.  There are also common analysis techniques and theoretical tools behind the identification of models and their relation to the native interactions in the quantum hardware.  A successful quantum simulator program will streamline the use of such common tools by establishing standards and encouraging the sharing of resources between all scientific groups, including software and theoretical computational tools, schematics for control systems used across many architectures, and the practical details of building quantum simulators on specific architectures that often prove surprisingly useful on different platforms.

Operating different platforms requires different equipment.  For example, trapped atoms or ions and superconducting quantum circuits are, at first sight, quite distinct. However, they have directly benefited from each other, for example in applying concepts of quantum trajectories from quantum optics~\cite{plenio1998quantum} to superconducting circuits~\cite{murch2013observing,quantum_jump}.
In fact, experimental technologies, specific techniques, physical models, system Hamiltonians, and quantum operations can all be very similar.  Dedicated funding support for establishing and promoting these interdisciplinary collaborations will have a high impact. In particular, to create the necessary synergy among laboratories, we envision a mixture of two approaches: vertical and horizontal.  In the vertically integrated approach, a single experimental quantum simulator platform is implemented by a team of PIs with different expertise. In the horizontally integrated approach, teams employ multiple platforms to study the same problem or model.

\subsection{Academic, National Laboratory, and Industry Cooperation and Mutual Benefit}

The success of these vertical and horizontal approaches requires continued maturation of key supporting sub-systems, including atomic sources, integrated optical systems, opto-mechanical assemblies, quantum-limited amplifiers, and high-speed field-programmable gate array (FPGA) control systems.  Given the significant challenges to maturation for quantum system engineering, we see a substantial opportunity to inspire and push for industrial development of these new tools for a broad user community through strategic and deliberately coordinated partnerships between academic, national lab, and industrial developers.

Collaborations between universities, national labs, and industry to improve existing technologies will also be vital to accelerating the development of quantum simulators as deployable devices. Advancements on this front can be spurred by supporting academic access to national lab and industrial systems. Dedicated or priority access at multiple levels (e.g., high-level software and API and low-level qubit and control hardware) will help enable researchers to understand the entire system and develop new protocols that will lead to improved and larger-scale quantum simulations. All parties can also provide open-source code building features that the community identifies as being beneficial for the advancement of their work.

We therefore envision a student- and postdoc- exchange program that will encourage cross-pollination of ideas at multiple levels.  First, we emphasize the need for direct interactions between researchers in experiment and theory. Theory students or postdocs will spend time at different institutions collaborating with  experimentalists in their labs and vice-versa. Second, students or postdocs from one architecture will visit labs from another to share common strategies that exist between different platforms.  Third, we suggest fellowships and other funding opportunities be created for students and post-docs to work in academia--national lab--industry partnerships.

It is imperative and timely to prepare students with a broad, cross-cutting educational experience spanning architectures---e.g., neutral atoms and superconducting quantum circuits---and sectors, e.g., academia and national labs. This approach will best position the future work-force to develop and deploy the prototypes needed to solve the scientifically pressing questions described in Sec.~\ref{sec:fundamental} and the practical problems highlighted in Sec.~\ref{sec:programmability}.

\subsection{National Quantum Repository and Information Exchange Hub}

Technical expertise in academia is typically siloed in individual groups and can be  difficult to share due to a lack of documentation, software robustness, and instrument compatibility. Strong collaborations and greater exchanges of technical expertise between research groups and experimental and theory programs, encouraged by dedicated funding, will strongly facilitate progress. For larger scale exchanges, we recommend establishing a national quantum repository and information exchange hub in the form of an internet resource for technical information and common basic subsystems.  An organized, searchable repository of schematics, technical drawings, and publications similar to the LIGO portal (\url{https://dcc.ligo.org/}) will be maintained as part of the quantum hub.

A key challenge to this plan is that physics research groups are typically not motivated to develop robust and well documented hardware solutions. A potential workaround to this problem is to fund engineering and computer science student projects  (e.g., as a supplement to a funded project) to design and benchmark a modular hardware or software component. Funding such (possibly co-advised) projects will provide a pathway to transferring critical engineering expertise into groups working on quantum simulator research. Information shared on the repository will also help identify current and future engineering needs for quantum simulators, which will help nucleate these student projects and activities.

The hub will also provide a centralized location to curate and promote existing open-source software tools, similar to the highly successful nanoHUB (\url{https://nanohub.org/}).  Funding agencies can incentivize and encourage groups to post capabilities to this hub. Discussion forums will also provide a centralized online communication network, allowing individuals from across physics, computer science, and engineering to communicate and grow collaborations involving expertise from different areas. Forum discussions will cover capabilities of experimental platforms and theoretical methods, motivating problems in physics and computer science and their connections to different hardware systems.

Creating such a forum and ensuring that it reaches critical mass (like the mathematics, computer science, and quantum information stack exchanges) will require a concerted effort and leadership from the community. Time and effort from experts in the field will be required to initiate and moderate discussions and establish a tradition of collaboration and communication. Resources will be required to enable principal investigators, postdocs, senior students, and dedicated engineers to devote time to standing up and developing the hub so that it can grow into a valuable tool for fostering collaboration, communication, and connections beyond a select few groups across the quantum simulator community.

We note that the success of the hub to accelerate the development of quantum technology hardware will depend critically on sustained and centralized support. This challenge provides an opportunity for the community to work with permanent partners such as national labs to establish such a critical resource.

In addition to a virtual hub, other approaches to a quantum simulator hub were discussed by workshop participants. Inspired by the LIGO-style focused collaborative effort, and motivated by potentially strong opportunities to bring revolutionary impacts to science and technology, a quantum simulator hub could include experimental scientists, engineers, theoretical physicists, and computer scientists all working together toward selected grand challenge goals. Such a Hub could utilize several complimentary platforms (both advanced and emerging) aiming to address key scientific and engineering challenges to produce practically useful simulators for scientific application. It would complement individual investigators' efforts and could include robust community outreach effort to enable access to most advanced platforms and to disseminate technology. It is also in the spirit of the National Quantum Initiative that a quantum simulator hub of this form would include strong partnerships with industry and national labs.

\section{Conclusions}
\label{sec:conclusions}

Despite advances in high-performance conventional computing, machine learning, and artificial intelligence, simulations on classical hardware remain unable to address many key scientific problems.  Universal digital quantum computers have been proposed to solve many of these outstanding problems, with very exciting recent progress toward quantum advantage and supremacy~\cite{arute2019quantum}.  However, achieving a sufficient number of well-controlled error-corrected qubits of large enough quantum volume~\cite{bishop2017quantum} to broadly realize impactful use-cases---i.e., the ``killer app''---is likely decades away.  An alternative, non-universal approach for certain applications are \emph{quantum simulators}. These devices are already being built on a wide variety of architectures (including atomic, molecular, optical, and solid-state systems), consist of up to millions of quantum elements, and range from highly optimized special-purpose devices to flexible and fully programmable machines.

Quantum simulators offer extraordinary opportunities for applications realizable on a 2--5 year time scale.  These include optimizing the performance of quantum materials, solving hard quantum chemistry problems related to reaction rates and catalysis, designing quantum devices to develop a fundamental understanding of  transport, modeling the quantum internet, and unlocking the potential quantum nature of photosynthesis.  Quantum simulators also provide a platform for resolving fundamental theoretical questions in cosmology, particle physics, gravity, and quantum thermodynamics.  Challenges to realizing this potential include scalability and complexity, state preparation and control, validation, verification, and error correction and mitigation.

To overcome these challenges, it is the position of the community that investment in a national quantum simulator program is a high priority. Such a program should consist of two main pillars: (1) \textbf{Early Prototype Quantum Simulators} will support the creation of quantum simulator prototypes usable by the broader scientific community, and (2) \textbf{New, Emerging Quantum Simulators} will support fundamental research carried out by a novel blend of multi-investigator, multi-disciplinary collaborations, including a quantum hub internet-based community resource for quantum simulator software, hardware, and education.  An integral part of pillars (1) and (2) are partnerships between university researchers, national labs, and industry.

The resulting goals for accelerating progress in quantum simulators resonate with the goals of two recent related NSF Accelerator workshops -- ``Quantum Interconnects'' and ``Quantum Computers.''

Any subjective views or opinions that might be expressed in the paper do not necessarily represent the views of the U.S. Department of Commerce, the U.S. Department of Energy, or the United States Government.

\noindent \textbf{Acknowledgments}: This material is based upon work supported by the National Science Foundation under Grant No. OIA-1945947.


\end{document}